\documentclass[twocolumn,showpacs,preprintnumbers,amsmath,amssymb]{revtex4}

\usepackage{graphicx}
\usepackage{dcolumn}
\usepackage{bm}
\usepackage{dsfont}

\newcommand{\eins}{\mathds{1}} 

\def\cyp{a}
\def\mainz{b}
\def\mit{c}
\def\athens{d}

\newcommand{\be}{\begin{equation}}
\newcommand{\ee}{\end{equation}}
\newcommand{\beq}{\begin{eqnarray}}
\newcommand{\eeq}{\end{eqnarray}}

\begin{document}


\title{$\Delta$-baryon electromagnetic form factors in lattice QCD}

\author{C. Alexandrou$^{(\cyp)}$, T. Korzec$^{(\cyp)}$, G. Koutsou$^{(\cyp)}$, Th. Leontiou$^{(\cyp)}$, C. Lorc\'e$^{(\mainz)}$, J. W. Negele$^{(\mit)}$,  V. Pascalutsa$^{(\mainz)}$,  A. Tsapalis$^{(\athens)}$, M. Vanderhaeghen$^{(\mainz)}$ }
\affiliation{
 {$^{(\cyp)}$ Department of Physics, University of Cyprus, P.O. Box 20537, 1678 Nicosia, Cyprus}\\ 
{  $^{(\mit)}$ Center for Theoretical Physics, 
Laboratory for
Nuclear Science and Department of Physics, Massachusetts Institute of
Technology, Cambridge, Massachusetts 02139, U.S.A.}\\
{$^{(\mainz)}$ Institut f\"ur Kernphysik, Johannes Gutenberg-Universit\"at, D-55099 Mainz, Germany }\\
{$^{(\athens)}$ Institute of Accelerating Systems and Applications, University of Athens, Athens} 
}


\begin{abstract}
We develop techniques  to calculate the four $\Delta$ electromagnetic form factors 
using lattice QCD,  with particular emphasis on the sub-dominant electric
quadrupole form factor that probes deformation of the $\Delta$. 
Results are  presented for pion masses down to approximately 350~MeV for three cases:
quenched QCD, two flavors of dynamical Wilson quarks, and three flavors of quarks described by a mixed action combining domain wall valence quarks and dynamical staggered sea quarks. 
The magnetic moment of the $\Delta$ is chirally extrapolated to the physical point 
and the $\Delta$ charge
density distributions are discussed. 
\end{abstract}

\maketitle

Lattice Quantum Chromodynamics (QCD) provides a well-defined framework to directly calculate hadron form factors from
the fundamental theory of  strong interactions. 
Form factors characterize
the internal structure of hadrons, including their magnetic moment, their size, and their charge
density distribution. Since the $\Delta(1232)$ decays strongly,
experiments~\cite{Kotulla:2002cg,LopezCastro:2000cv} to measure its form factors are harder and yield 
less precise results than for  nucleons~\cite{Arrington:2006zm,Perdrisat:2006hj}.
In this work, we compute $\Delta$ form factors using lattice QCD
more accurately than can be currently obtained from  experiment.

A primary motivation for this work is to understand the role of deformation in baryon structure:  whether any of the low-lying baryons have deformed intrinsic states and if so, why. Thus, a major achievement of this work is the development of lattice methods with sufficient precision to show, for the first time, that the 
electric quadrupole form factor is non-zero and hence the $\Delta$ has a non-vanishing quadrupole moment and an  associated deformed shape.  Unlike the $\Delta$, the spin-1/2 nucleon cannot have a quadrupole moment, so the experiment of choice to explore its deformation has been measurement of the nucleon to $\Delta$ electric and Coulomb quadrupole transition form factors. 
Major experiments~\cite{Mertz:2001,Joo:2001,Sparveris:2005} have shown that these transition form factors  are indeed
non-zero,  confirming the presence of deformation in either the 
nucleon, $\Delta$, or both~\cite{Papanicolas:2006athens,Papanicolas:2003}, and  lattice QCD
yields comparable non-zero results~\cite{Alexandrou:EM_N2D2004prl,Alexandrou:2007dt}. 
Our new calculation of the $\Delta$ quadrupole form factor, coupled with the nucleon to $\Delta$ transition form factors,
should in turn shed light on the deformation of the nucleon.

In order to evaluate the $\Delta$ electromagnetic (EM) form factors to the required accuracy,  we  
isolate the two dominant form factors and the sub-dominant electric quadrupole form factor. This is particularly crucial for the
latter since it can be extracted with greater precision, although it
increases the computational cost. 
Our techniques are first tested in quenched QCD~\cite{Alexandrou:2007we}.
 We then
calculate form factors
using two degenerate flavors of dynamical
Wilson fermions, denoted by $N_F=2$, with pion masses 
in the range of 700~MeV to 380 MeV~\cite{Orth:2005kq,Urbach:2005ji}. Finally,
we use a mixed action with chirally symmetric domain wall valence quarks  and staggered sea quarks
with two degenerate light flavors and one strange flavor~\cite{Orginos:1999cr}, denoted by $N_F=2+1$,  at
a pion mass of 353~MeV. 
Using the results obtained with
dynamical quarks, we extrapolate
the magnetic moment to the physical point. We extract the quark charge 
distributions in the $\Delta$, and discuss their
quadrupole moment.

The $\Delta$  matrix element $\langle \Delta(p_f,s_f) | j_{\rm EM}^\mu | \Delta(p_i,s_i)\rangle$, where $j_{\rm EM}^\mu$ is the electromagnetic current,
can be parametrized in terms of four multipole form factors that 
depend only on the momentum transfer $q^2 \equiv -Q^2 =(p_f-p_i)^2$~\cite{Nozawa:1990gt}. 
The decomposition for the
on shell $\gamma^* \Delta\Delta$ matrix
element is given by  
\beq
 &\>&\langle \Delta(p_f,s_f) | j_{\rm EM}^\mu | \Delta(p_i,s_i)\rangle={\cal A}\>\>\bar{u}_\sigma(p_f,s_f){\cal O}^{\sigma \mu \tau} u_\tau(p_i,s_i) \nonumber \\
&\>&{\cal O}^{\sigma \mu \tau}=-g^{\sigma \tau}\biggl[a_1(q^2) \gamma^\mu +\frac{a_2(q^2)}{2m_\Delta} \left(p_f^\mu + p_i^\mu\right)\biggr] \nonumber \\
&\>&\hspace*{1.3cm}-\frac{q^\sigma q^\tau}{4m_\Delta^2}\biggl[c_1(q^2)\gamma^\mu + \frac{c_2(q^2)}{2m_\Delta}\left(p_f^\mu+p_i^\mu\right)\biggr] ,
\eeq
 where  $a_1(q^2)$, $a_2(q^2)$, $c_1(q^2)$, and $c_2(q^2)$ are  known linear combinations of   
the electric charge form factor $G_{E0}(q^2)$,
the magnetic dipole form factor $G_{M1}(q^2)$, the electric quadrupole form factor $G_{E2}(q^2)$,
and the magnetic octupole form factor $G_{M3}(q^2)$~\cite{Leinweber:1992hy}, 
and $\cal{A}$ is  a known
factor depending on the normalization of hadron states.
These form factors can be extracted from
correlation functions calculated in lattice QCD~\cite{Leinweber:1992hy}.
We calculate in Euclidean time the  two- and three-point correlation functions in a frame where the final state $\Delta$ is at rest:  
\begin{eqnarray}
 G(t,\vec q) &=& \sum_{\vec x_f}\sum_{j=1}^3 e^{-i\vec x_f \cdot \vec q}\, 
       \Gamma^4_{\alpha\beta}   \langle J_{j\beta}(x_f)  \overline J_{j\alpha}(0) \rangle\nonumber\\
\hspace*{-0.5cm} G^{\ \mu}_{\sigma\ \tau}(\Gamma^\nu,t,\vec q) &=& \sum_{\vec x_f \vec x} e^{i\vec x \cdot \vec q} 
       \Gamma^\nu_{\alpha\beta} \langle J_{\sigma\beta}(x_f) j^\mu(x) \overline J_{\tau\alpha}(0) \rangle , 
\end{eqnarray}
where $j^\mu$ is the electromagnetic current on the lattice, $J$ and  $\overline J$ are 
the $\Delta^+$ interpolating fields constructed from smeared quarks~\cite{Alexandrou:2007we}, 
$\Gamma^4=\frac{1}{4}(\eins+\gamma^4)$, and $\Gamma^k=i\Gamma^4\gamma^5\gamma^k$. 
The form factors can then be extracted from ratios of three- and two-point functions
in which unknown normalization constants and the leading time dependence cancel
\begin{equation}
       R_{\sigma\ \tau}^{\ \mu} = \frac{G_{\sigma\ \tau}^{\ \mu}(\Gamma,t,\vec q)}{G(t_f,\vec 0)}\ 
				         \sqrt{\frac{G(t_f-t,\vec p_i)G(t,\vec 0  )G(t_f,\vec 0  )}
					            {G(t_f-t,\vec 0  )G(t,\vec p_i)G(t_f,\vec p_i)}} \, .
\end{equation}
For sufficiently large $t_f-t$ and $t-t_i$, this ratio exhibits a plateau
$R(\Gamma,t,\vec q) \to \Pi(\Gamma,\vec q)$, from which  
the form factors are extracted, and we use the particular combinations 
\begin{eqnarray}
 &\>& \sum_{k=1}^3                    \Pi_{k\ k}^{\ \mu}(\Gamma^4,\vec q) = K_1\ G_{E0}(Q^2) + K_2\ G_{E2}(Q^2)\label{type3}\\
 &\>& \sum_{j,k,l=1}^3 \epsilon_{jkl} \Pi_{j\ k}^{\ \mu}(\Gamma^4,\vec q) = K_3\ G_{M1}(Q^2) \label{type2}\\
  &\>& \sum_{j,k,l}^3   \epsilon_{jkl} \Pi_{j\ k}^{\ 4}(\Gamma^j,\vec q)   = K_4\ G_{E2}(Q^2) \, . \label{type4}
\end{eqnarray}
The connected part of each combination of three-point functions
 can be calculated  efficiently using the method of sequential
 inversions~\cite{Dolgov:2002zm}. 
 At present,
 it is not yet computationally feasible to calculate 
the small corrections arising from disconnected diagrams. 
The known kinematical
coefficients $K_1, K_2, K_3, K_4$ are functions of the $\Delta$ mass and
energy as well as of $\mu$ and $\vec q$. The combinations above
are chosen such that all possible directions of $\mu$ and $\vec q$ 
contribute symmetrically to the form factors at a given $Q^2$~\cite{Marc}. 
The over-constrained system of Eqs.~(\ref{type3}-\ref{type4}) is solved by a least-squares analysis, and $G_{E2}(Q^2) $ can also be isolated separately from 
Eq.~(\ref{type4}).

The details of the simulations are 
summarized in Table~\ref{bigtab}.
In each case, the separation between the final and
initial time is $t_f-t_i \gtrsim 1\,{\rm fm}$ and Gaussian smearing is
applied to both source and sink to produce adequate plateaus by suppressing contamination from higher
states having the  quantum numbers of the $\Delta(1232)$. 
For the mixed-action calculation, the 
domain-wall valence quark mass was chosen  to reproduce
the lightest pion mass obtained using $N_F=2+1$
improved staggered quarks~\cite{Edwards:2005ym, Marc}.

\begin{table}[h]
\caption{\label{bigtab}Lattice parameters and results.  $N_{\rm conf}$ denotes the
number of lattice configurations, $\sqrt{\langle r^2\rangle}$ gives the charge radius,  $\mu_{\Delta^+}$ is the $\Delta^+$ magnetic moment in nuclear
magnetons and $Q^\Delta_{\frac{3}{2}}$ is the $\Delta^+$ quadrupole moment. }
\begin{ruledtabular}
\begin{tabular}{ c c c c c c}
 $N_{\rm conf}$ & $m_\pi$ [GeV] & $m_\Delta$ [GeV] & $\sqrt{\langle r^2\rangle}$ [fm] & $\mu_{\Delta^+}$ [$\mu_N$]& $Q^\Delta_{\frac{3}{2}}$ \\
\hline
\multicolumn{6}{c}{Quenched Wilson, $32^3\times 64$, $a=0.092$~fm}\\
\hline
 200       & 0.563(4)      & 1.470(15)        &    0.6147(66)       &   1.720(42)   &  0.96(12)  \\
 200       & 0.490(4)      & 1.425(16)        &    0.6329(76)       &   1.763(51)   &  0.91(15)        \\
 200       & 0.411(4)      & 1.382(19)        &    0.6516(87)       &   1.811(69)   &  0.83(21)           \\
\hline
\multicolumn{6}{c}{$N_F=2$ Wilson, $24^3\times 40 (32$ for lightest pion), $a=0.077$~fm 
}\\
\hline
 185       & 0.691(8)      & 1.687(15)        &  0.5279(61)       &   1.462(45) & 0.80(21)            \\
157       & 0.509(8)       & 1.559(19)        &  0.594(10)        &   1.642(81) &  0.41(45)       \\
 200       & 0.384(8)      & 1.395(18)        &  0.611(17)        &   1.58(11)  & 0.46(35)      \\
\hline
\multicolumn{6}{c}{$N_F=2+1$, Mixed action, $28^3\times 64$, $a=0.124$~fm \cite{Aubin:2004wf}}\\
\hline
 300       & 0.353(2)      & 1.533(27)        &  0.641(22)          &    1.91(16)   & 0.74(68)          \\
\end{tabular}
\end{ruledtabular}
\vspace*{-0.5cm}
\end{table}

\begin{figure}[h]
\includegraphics[width=8cm,height=6cm]{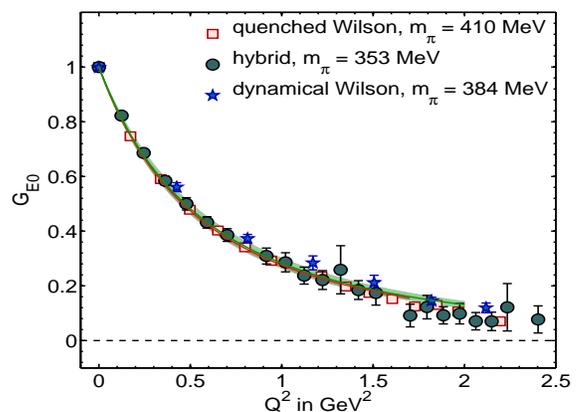}
\vspace*{-0.5cm}
\caption{The electric charge form factor versus $Q^2$. The green (red) 
line and error band show 
a dipole fit to the  mixed action (quenched ) results.
}\label{ge0fig}
\vspace*{-0.5cm}
\end{figure}

The results for $G_{E0}(Q^2)$ are shown
in Fig.~\ref{ge0fig} as a function of  $Q^2$ at the lightest
pion mass for each of the three actions.
For Wilson fermions, we use the conserved lattice current requiring no renormalization.  The local current is used for the mixed action, and the
renormalization constant,  $Z_V=1.0992(32)$, is determined
by  the condition that $G_{E0}(0)$  equals the charge of the $\Delta$ in units of $e$.
As can be seen, all three calculations yield consistent results.
The momentum dependence of the charge form factor is 
described well by a dipole form
$  G_{E0}(Q^2) = 1/\left(1+ \frac{Q^2}{\Lambda_{E0}^2}\right)^2 \,$ .
To compare the slopes at $Q^2=0$, we follow convention and show in table~\ref{bigtab} the 
 so-called ``rms radius''~\cite{Leinweber:1992hy} 
$  \left\langle r^2 \right\rangle = -6 \left. \frac{d}{dQ^2} G_{E0}(Q^2)\right|_{Q^2=0} \, .$
%

\begin{figure}[h]
\includegraphics[width=8cm,height=6cm]{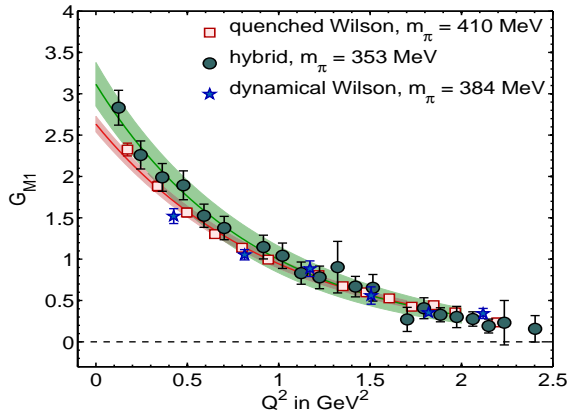}
\vspace*{-0.5cm}
\caption{The magnetic dipole form factor. The green (red) line and error band show
an exponential fit to the  mixed action (quenched )results.}
\label{gm1fig}
\end{figure}
\begin{figure}[h]
\includegraphics[width=7cm,height=5cm]{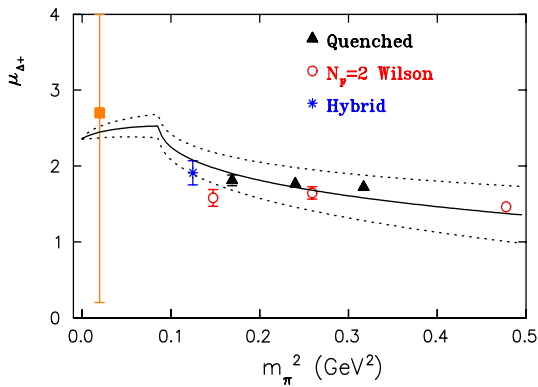}
\caption{The magnetic dipole moment in nuclear magnetons. The value at the physical pion mass (filled square) is shown with  statistical and systematic
errors~\cite{Kotulla:2002cg}.  The solid and dashed curves  show the chiral extrapolation and  theoretical error estimate~\cite{Pascalutsa:2004je} .}
\label{fig:mu}
\vspace*{-0.5cm}
\end{figure}

The momentum dependence of $G_{M1}(Q^2)$ is displayed in Fig.~\ref{gm1fig}.
To extract the magnetic moment, 
an extrapolation to 
zero momentum transfer is necessary. Both an exponential
form, $G_{M1} e^{-Q^2/\Lambda_{M1}^2}$, and a dipole describe the $Q^2$-dependence well, and we adopt the exponential form because of its faster decay at
large $Q^2$, in accord with perturbative arguments. 
The larger spatial volume for the quenched and mixed action cases 
yields smaller and more densely spaced values of the lattice momenta 
and correspondingly more precise determination of the form factor 
than for the smaller volume used with dynamical Wilson fermions.
In Fig.~\ref{gm1fig}, we show the best exponential fit and error band for the mixed action and quenched results.
As can be seen, results in the quenched theory and
for $N_F=2$ Wilson fermions are within the error band.   
The magnetic moment in natural units is
given by $\mu_\Delta=G_{M1}(0)e/(2m_\Delta)$, where 
$m_\Delta$ is the $\Delta$  mass measured on the lattice
and  $G_{M1}(0)$ is from  the exponential fits. 
In Table~\ref{bigtab} we give the values of the ${\Delta^+}$ magnetic moment
in nuclear magnetons $e/(2 M_N)$, with $M_N$ the physical nucleon mass. 
The magnetic moments of the $\Delta^+$ and $\Delta^{++}$ are accessible to 
experiments~\cite{Kotulla:2002cg,LopezCastro:2000cv}, which presently suffer from large uncertainties. 
The magnetic moment as a function of $m_\pi^2$ is shown in Fig.~\ref{fig:mu}, together with
a chiral extrapolation to the physical point~\cite{Pascalutsa:2004je}, which lies
  within the broad error band $\mu_{\Delta^{+}}= 2.7^{+1.0}_{-1.3}(stat.)\pm 1.5 (syst.) \pm 3.0 (theory)\mu_N$~\cite{Kotulla:2002cg}.
The $\Delta$ moments using an approach similar to ours
 are calculated only in the  quenched approximation~\cite{Leinweber:1992hy,Zanotti:2004jy,Boinepalli:2006ng}.
Our magnetic moment results agree  with recent 
background field calculations using dynamical improved Wilson
fermions~\cite{Aubin:2008hz}, which supersede previous quenched background field results~\cite{Lee:2005ds}. 
The spatial length $L_s$ of our lattices
satisfies $L_s m_\pi> 4 $ in all cases 
except at the lightest pion mass with $N_F=2$ Wilson fermions, for which 
$L_s m_\pi =3.6$.  For that point,  the magnetic moment falls slightly below the error band, consistent with the fact that Ref.~\cite{Aubin:2008hz} shows that finite volume effects  decrease the magnetic moment.

\begin{figure}[h]
\includegraphics[width=7.5cm,height=5.5cm]{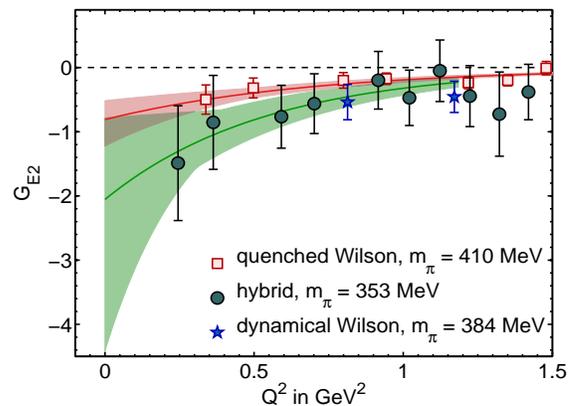}
\vspace*{-0.5cm}
\caption{The electric quadrupole form factor. The notation is the same as that in Fig.~\ref{ge0fig}.}\label{ge2fig}
\vspace*{-0.5cm}
\end{figure}

The electric quadrupole form factor is particularly interesting because
 it can be related to the shape of a hadron, and lattice calculations
 for each of the three actions are shown in Fig.~\ref{ge2fig} 
with exponential fits for the quenched and mixed action cases. 
Just as the electric form factor for a spin 1/2 nucleon can be 
expressed precisely as the transverse Fourier transform of the 
transverse quark charge density in the infinite momentum frame~\cite{Burkardt:2000za}, a  proper field-theoretic 
interpretation of the shape of the $\Delta(1232)$ can be 
obtained by considering the quark transverse
charge densities in this frame~\cite{Miller:2007uy,Carlson:2007xd,Carlson:2008zc}.
With respect to the direction of the average baryon momentum $P$, 
the transverse charge density in a spin-3/2 state with transverse 
polarization $s_\perp$ is defined as~:
\begin{eqnarray}
\rho^\Delta_{T \, s_\perp}(\vec b) 
&\equiv& \int \frac{d^2 \vec q_\perp}{(2 \pi)^2} \,
e^{- i \, \vec q_\perp \cdot \vec b} \, \frac{1}{2 P^+} \nonumber \\
&&\times \langle P^+, \frac{\vec q_\perp}{2}, s_\perp  
\,|\, J^+(0) \,|\, P^+, -\frac{\vec q_\perp}{2}, s_\perp   \rangle, 
\label{eq:dens3}
\end{eqnarray}
where the photon transverse momentum $\vec q_\perp$ satisfies 
$\vec q_\perp^{\, 2} = Q^2$,
$J^+ \equiv J^0 +J^3$,
and $\vec b$ specifies the quark position  in the 
$xy$-plane  relative to  the $\Delta$ center of mass. 
Choosing the $\Delta$ transverse spin vector along the $x$-axis, the 
quadrupole moment of this two-dimensional charge distribution is
defined as~\cite{Marc}: 
\begin{eqnarray}
Q^\Delta_{s_\perp} \equiv e \int d^2 \vec b \, (b_x^2 - b_y^2) \, 
\rho^\Delta_{T \,  s_\perp}(\vec b). 
\end{eqnarray}
In terms of the $\Delta$ EM form factors~\cite{Marc}~,
\begin{eqnarray}
Q^\Delta_{\frac{3}{2}} = \frac{1}{2} 
\left\{ 2 \left[ G_{M1}(0) - 3 e_\Delta \right] + 
\left[ G_{E2}(0) + 3 e_\Delta \right] \right\} 
\frac{e}{M_\Delta^2}.
\label{eq:quadrup}
\end{eqnarray}
The term proportional 
to $[G_{M1}(0) -3 e_\Delta]$ is an electric quadrupole moment 
induced in the moving frame due to the magnetic dipole moment.   
For a spin-3/2 particle without internal structure, 
$G_{M1}(0) = 3 e_\Delta$, 
$G_{E2}(0) = -3 e_\Delta$~\cite{Deser:2000dz,Marc}, 
and the quadrupole moment of the 
transverse charge density vanishes.  
Hence $Q^\Delta_{s_\perp}$, and thus the deformation of the two dimensional
transverse charge 
density, is only sensitive to the {\it anomalous} parts 
of the spin-3/2 magnetic dipole and electric quadrupole moments, 
and vanishes for a particle without internal structure. The analogous property holds for a spin-1 particle~\cite{Carlson:2008zc}, 
indicating the generality of this description in terms of transverse
densities.

\begin{figure}[h]
\begin{center}
\includegraphics[width=5.5cm,height=5.5cm]{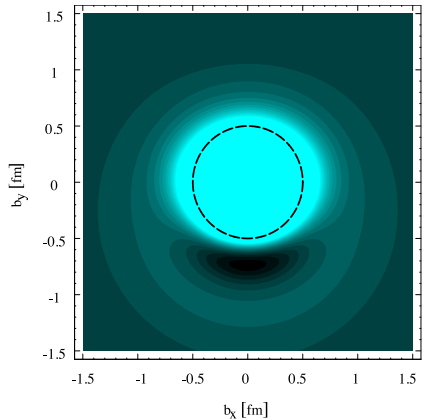}
\end{center}
\vspace*{-0.5cm}
\caption{Quark transverse charge density in a 
{\it $\Delta^+$} polarized along the $x$-axis, with $s_\perp = +3/2$. 
The light (dark) regions correspond with
the largest (smallest) values of the density.} 
\label{fig:deltatrans}
\vspace*{-0.3cm}
\end{figure}

Fig.~\ref{fig:deltatrans} shows the transverse density  
$\rho^\Delta_{T \, s_\perp}$ 
for a $\Delta^+$ with transverse spin $s_\perp = +3/2$ calculated from the fit to the quenched Wilson lattice results for the $\Delta$  
form factors (which has the smallest statistical errors of the three calculations). 
It is seen that the $\Delta^+$ quark charge density is elongated 
along the axis of the spin (prolate).   
This prolate deformation is robust in the sense that the values for $Q^\Delta_{\frac{3}{2}}$ obtained from  
Eq.~(\ref{eq:quadrup}) and given in Table~\ref{bigtab} are all consistently positive.
%

In the case of the magnetic octupole form factor~\cite{Marc}, which is  related to the magnetic octupole moment
${\cal O}_\Delta=G_{M3}(0)e/ 2 m_\Delta^3$, our statistics are insufficient to distinguish the result from zero.


In summary, a formalism 
for the accurate evaluation of the $\Delta$ electromagnetic form factors
as  functions of $q^2$ has been developed and used in quenched QCD 
and full QCD with $N_F$ = 2 and 2+1 flavors. 
The charge radius and magnetic dipole moment were determined as a function 
of $m_\pi^2$ and the dipole moment was chirally extrapolated
to the physical point. 
The electric quadrupole form factor was evaluated for the first time
with sufficient accuracy to distinguish it from zero. The lattice calculations show that 
the quark density in a $\Delta^+$ of transverse spin projection +3/2 is
elongated along the spin axis.

\centerline{\bf Acknowledgments}
This work is
supported in part by the Cyprus Research Promotion Foundation (RPF) under contract $\Pi$ENEK/ENI$\Sigma$X/0505-39, the  EU Integrated Infrastructure Initiative
Hadron Physics (I3HP) under contract RII3-CT-2004-506078 and 
the
U.S. Department of Energy (D.O.E.) Office of Nuclear Physics under contracts
DE-FG02-94ER40818 and DE-FG02-04ER41302.
This research used computational resources provided by RFP under contract EPYAN/0506/08, 
 the National Energy Research Scientific Computing  Center 
supported by the Office of Science of the U.S. Department of Energy under Contract DE-AC03-76SF00098
and 
the MIT Blue Gene computer 
under grant DE-FG02-05ER25681.
Dynamical staggered quark configurations  and forward domain wall quark propagators were provided by the MILC and LHPC collaborations respectively. 

\bibliography{dd}

\begin{thebibliography}{31}
\expandafter\ifx\csname natexlab\endcsname\relax\def\natexlab#1{#1}\fi
\expandafter\ifx\csname bibnamefont\endcsname\relax
  \def\bibnamefont#1{#1}\fi
\expandafter\ifx\csname bibfnamefont\endcsname\relax
  \def\bibfnamefont#1{#1}\fi
\expandafter\ifx\csname citenamefont\endcsname\relax
  \def\citenamefont#1{#1}\fi
\expandafter\ifx\csname url\endcsname\relax
  \def\url#1{\texttt{#1}}\fi
\expandafter\ifx\csname urlprefix\endcsname\relax\def\urlprefix{URL }\fi
\providecommand{\bibinfo}[2]{#2}
\providecommand{\eprint}[2][]{\url{#2}}

\bibitem[{\citenamefont{Kotulla et~al.}(2002)}]{Kotulla:2002cg}
\bibinfo{author}{\bibfnamefont{M.}~\bibnamefont{Kotulla}} \bibnamefont{et~al.},
  \bibinfo{journal}{Phys. Rev. Lett.} \textbf{\bibinfo{volume}{89}},
  \bibinfo{pages}{272001} (\bibinfo{year}{2002}).

\bibitem[{\citenamefont{Lopez~Castro and Mariano}(2001)}]{LopezCastro:2000cv}
\bibinfo{author}{\bibfnamefont{G.}~\bibnamefont{Lopez~Castro}}
  \bibnamefont{and} \bibinfo{author}{\bibfnamefont{A.}~\bibnamefont{Mariano}},
  \bibinfo{journal}{Phys. Lett.} \textbf{\bibinfo{volume}{B517}},
  \bibinfo{pages}{339} (\bibinfo{year}{2001}).

\bibitem[{\citenamefont{Arrington et~al.}(2007)\citenamefont{Arrington,
  Roberts, and Zanotti}}]{Arrington:2006zm}
\bibinfo{author}{\bibfnamefont{J.}~\bibnamefont{Arrington}},
  \bibinfo{author}{\bibfnamefont{C.~D.} \bibnamefont{Roberts}},
  \bibnamefont{and} \bibinfo{author}{\bibfnamefont{J.~M.}
  \bibnamefont{Zanotti}}, \bibinfo{journal}{J. Phys.}
  \textbf{\bibinfo{volume}{G34}}, \bibinfo{pages}{S23} (\bibinfo{year}{2007}),
  \eprint{nucl-th/0611050}.

\bibitem[{\citenamefont{Perdrisat et~al.}(2007)\citenamefont{Perdrisat,
  Punjabi, and Vanderhaeghen}}]{Perdrisat:2006hj}
\bibinfo{author}{\bibfnamefont{C.~F.} \bibnamefont{Perdrisat}},
  \bibinfo{author}{\bibfnamefont{V.}~\bibnamefont{Punjabi}}, \bibnamefont{and}
  \bibinfo{author}{\bibfnamefont{M.}~\bibnamefont{Vanderhaeghen}},
  \bibinfo{journal}{Prog. Part. Nucl. Phys.} \textbf{\bibinfo{volume}{59}},
  \bibinfo{pages}{694} (\bibinfo{year}{2007}).

\bibitem[{\citenamefont{Mertz et~al.}(2001)}]{Mertz:2001}
\bibinfo{author}{\bibfnamefont{C.}~\bibnamefont{Mertz}} \bibnamefont{et~al.}
  (\bibinfo{collaboration}{OOPS}), \bibinfo{journal}{Phys. Rev. Lett.}
  \textbf{\bibinfo{volume}{86}}, \bibinfo{pages}{2963} (\bibinfo{year}{2001}).

\bibitem[{\citenamefont{Joo et~al.}(2002)}]{Joo:2001}
\bibinfo{author}{\bibfnamefont{K.}~\bibnamefont{Joo}} \bibnamefont{et~al.}
  (\bibinfo{collaboration}{CLAS}), \bibinfo{journal}{Phys. Rev. Lett.}
  \textbf{\bibinfo{volume}{88}}, \bibinfo{pages}{122001}
  (\bibinfo{year}{2002}).

\bibitem[{\citenamefont{Sparveris et~al.}(2005)}]{Sparveris:2005}
\bibinfo{author}{\bibfnamefont{N.~F.} \bibnamefont{Sparveris}}
  \bibnamefont{et~al.}, \bibinfo{journal}{Phys. Rev. Lett.}
  \textbf{\bibinfo{volume}{94}}, \bibinfo{pages}{022003}
  (\bibinfo{year}{2005}).

\bibitem[{\citenamefont{Papanicolas and
  Bernstein}(2007)}]{Papanicolas:2006athens}
\bibinfo{author}{\bibfnamefont{C.~N.} \bibnamefont{Papanicolas}}
  \bibnamefont{and} \bibinfo{author}{\bibfnamefont{A.~M.}
  \bibnamefont{Bernstein}}, \bibinfo{journal}{AIP Conference Proceedings}
  \textbf{\bibinfo{volume}{104}}, \bibinfo{pages}{1} (\bibinfo{year}{2007}).

\bibitem[{\citenamefont{Papanicolas}(2003)}]{Papanicolas:2003}
\bibinfo{author}{\bibfnamefont{C.~N.} \bibnamefont{Papanicolas}},
  \bibinfo{journal}{Eur. Phys. J.} \textbf{\bibinfo{volume}{A18}},
  \bibinfo{pages}{141} (\bibinfo{year}{2003}).

\bibitem[{\citenamefont{Alexandrou et~al.}(2005)}]{Alexandrou:EM_N2D2004prl}
\bibinfo{author}{\bibfnamefont{C.}~\bibnamefont{Alexandrou}}
  \bibnamefont{et~al.}, \bibinfo{journal}{Phys. Rev. Lett.}
  \textbf{\bibinfo{volume}{94}}, \bibinfo{pages}{021601}
  (\bibinfo{year}{2005}).

\bibitem[{\citenamefont{Alexandrou et~al.}(2008)}]{Alexandrou:2007dt}
\bibinfo{author}{\bibfnamefont{C.}~\bibnamefont{Alexandrou}}
  \bibnamefont{et~al.}, \bibinfo{journal}{Phys. Rev.}
  \textbf{\bibinfo{volume}{D77}}, \bibinfo{pages}{085012}
  (\bibinfo{year}{2008}).

\bibitem[{\citenamefont{Alexandrou et~al.}(2007)\citenamefont{Alexandrou,
  Korzec, Leontiou, Negele, and Tsapalis}}]{Alexandrou:2007we}
\bibinfo{author}{\bibfnamefont{C.}~\bibnamefont{Alexandrou}},
  \bibinfo{author}{\bibfnamefont{T.}~\bibnamefont{Korzec}},
  \bibinfo{author}{\bibfnamefont{T.}~\bibnamefont{Leontiou}},
  \bibinfo{author}{\bibfnamefont{J.~W.} \bibnamefont{Negele}},
  \bibnamefont{and} \bibinfo{author}{\bibfnamefont{A.}~\bibnamefont{Tsapalis}},
  \bibinfo{journal}{PoS} \textbf{\bibinfo{volume}{2007}}, \bibinfo{pages}{149}
  (\bibinfo{year}{2007}).

\bibitem[{\citenamefont{Orth et~al.}(2005)\citenamefont{Orth, Lippert, and
  Schilling}}]{Orth:2005kq}
\bibinfo{author}{\bibfnamefont{B.}~\bibnamefont{Orth}},
  \bibinfo{author}{\bibfnamefont{T.}~\bibnamefont{Lippert}}, \bibnamefont{and}
  \bibinfo{author}{\bibfnamefont{K.}~\bibnamefont{Schilling}},
  \bibinfo{journal}{Phys. Rev.} \textbf{\bibinfo{volume}{D72}},
  \bibinfo{pages}{014503} (\bibinfo{year}{2005}).

\bibitem[{\citenamefont{Urbach et~al.}(2006)\citenamefont{Urbach, Jansen,
  Shindler, and Wenger}}]{Urbach:2005ji}
\bibinfo{author}{\bibfnamefont{C.}~\bibnamefont{Urbach}},
  \bibinfo{author}{\bibfnamefont{K.}~\bibnamefont{Jansen}},
  \bibinfo{author}{\bibfnamefont{A.}~\bibnamefont{Shindler}}, \bibnamefont{and}
  \bibinfo{author}{\bibfnamefont{U.}~\bibnamefont{Wenger}},
  \bibinfo{journal}{Comput. Phys. Commun.} \textbf{\bibinfo{volume}{174}},
  \bibinfo{pages}{87} (\bibinfo{year}{2006}).

\bibitem[{\citenamefont{Orginos et~al.}(1999)\citenamefont{Orginos, Toussaint,
  and Sugar}}]{Orginos:1999cr}
\bibinfo{author}{\bibfnamefont{K.}~\bibnamefont{Orginos}},
  \bibinfo{author}{\bibfnamefont{D.}~\bibnamefont{Toussaint}},
  \bibnamefont{and} \bibinfo{author}{\bibfnamefont{R.~L.} \bibnamefont{Sugar}}
  (\bibinfo{collaboration}{MILC}), \bibinfo{journal}{Phys. Rev.}
  \textbf{\bibinfo{volume}{D60}}, \bibinfo{pages}{054503}
  (\bibinfo{year}{1999}).

\bibitem[{\citenamefont{Nozawa and Leinweber}(1990)}]{Nozawa:1990gt}
\bibinfo{author}{\bibfnamefont{S.}~\bibnamefont{Nozawa}} \bibnamefont{and}
  \bibinfo{author}{\bibfnamefont{D.~B.} \bibnamefont{Leinweber}},
  \bibinfo{journal}{Phys. Rev.} \textbf{\bibinfo{volume}{D42}},
  \bibinfo{pages}{3567} (\bibinfo{year}{1990}).

\bibitem[{\citenamefont{Leinweber et~al.}(1992)\citenamefont{Leinweber, Draper,
  and Woloshyn}}]{Leinweber:1992hy}
\bibinfo{author}{\bibfnamefont{D.~B.} \bibnamefont{Leinweber}},
  \bibinfo{author}{\bibfnamefont{T.}~\bibnamefont{Draper}}, \bibnamefont{and}
  \bibinfo{author}{\bibfnamefont{R.~M.} \bibnamefont{Woloshyn}},
  \bibinfo{journal}{Phys. Rev.} \textbf{\bibinfo{volume}{D46}},
  \bibinfo{pages}{3067} (\bibinfo{year}{1992}).

\bibitem[{\citenamefont{Dolgov et~al.}(2002)}]{Dolgov:2002zm}
\bibinfo{author}{\bibfnamefont{D.}~\bibnamefont{Dolgov}} \bibnamefont{et~al.}
  (\bibinfo{collaboration}{LHPC}), \bibinfo{journal}{Phys. Rev.}
  \textbf{\bibinfo{volume}{D66}}, \bibinfo{pages}{034506}
  (\bibinfo{year}{2002}).

\bibitem[{\citenamefont{Alexandrou et~al.}()}]{Marc}
\bibinfo{author}{\bibfnamefont{C.}~\bibnamefont{Alexandrou}}
  \bibnamefont{et~al.}, \bibinfo{note}{in preparation}.

\bibitem[{\citenamefont{Edwards et~al.}(2006)}]{Edwards:2005ym}
\bibinfo{author}{\bibfnamefont{R.~G.} \bibnamefont{Edwards}}
  \bibnamefont{et~al.} (\bibinfo{collaboration}{LHPC}), \bibinfo{journal}{Phys.
  Rev. Lett.} \textbf{\bibinfo{volume}{96}}, \bibinfo{pages}{052001}
  (\bibinfo{year}{2006}).

\bibitem[{\citenamefont{Aubin et~al.}(2004)}]{Aubin:2004wf}
\bibinfo{author}{\bibfnamefont{C.}~\bibnamefont{Aubin}} \bibnamefont{et~al.},
  \bibinfo{journal}{Phys. Rev.} \textbf{\bibinfo{volume}{D70}},
  \bibinfo{pages}{094505} (\bibinfo{year}{2004}).

\bibitem[{\citenamefont{Pascalutsa and
  Vanderhaeghen}(2005)}]{Pascalutsa:2004je}
\bibinfo{author}{\bibfnamefont{V.}~\bibnamefont{Pascalutsa}} \bibnamefont{and}
  \bibinfo{author}{\bibfnamefont{M.}~\bibnamefont{Vanderhaeghen}},
  \bibinfo{journal}{Phys. Rev. Lett.} \textbf{\bibinfo{volume}{94}},
  \bibinfo{pages}{102003} (\bibinfo{year}{2005}).

\bibitem[{\citenamefont{Boinepalli et~al.}(2006)}]{Boinepalli:2006ng}
\bibinfo{author}{\bibfnamefont{S.}~\bibnamefont{Boinepalli}}
  \bibnamefont{et~al.}, \bibinfo{journal}{PoS}
  \textbf{\bibinfo{volume}{LAT2006}}, \bibinfo{pages}{124}
  (\bibinfo{year}{2006}).

\bibitem[{\citenamefont{Zanotti et~al.}(2004)\citenamefont{Zanotti, Boinepalli,
  Leinweber, Williams, and Zhang}}]{Zanotti:2004jy}
\bibinfo{author}{\bibfnamefont{J.~M.} \bibnamefont{Zanotti}},
  \bibinfo{author}{\bibfnamefont{S.}~\bibnamefont{Boinepalli}},
  \bibinfo{author}{\bibfnamefont{D.~B.} \bibnamefont{Leinweber}},
  \bibinfo{author}{\bibfnamefont{A.~G.} \bibnamefont{Williams}},
  \bibnamefont{and} \bibinfo{author}{\bibfnamefont{J.~B.} \bibnamefont{Zhang}},
  \bibinfo{journal}{Nucl. Phys. Proc. Suppl.} \textbf{\bibinfo{volume}{128}},
  \bibinfo{pages}{233} (\bibinfo{year}{2004}).

\bibitem[{\citenamefont{Aubin et~al.}(2008)\citenamefont{Aubin, Orginos,
  Pascalutsa, and Vanderhaeghen}}]{Aubin:2008hz}
\bibinfo{author}{\bibfnamefont{C.}~\bibnamefont{Aubin}},
  \bibinfo{author}{\bibfnamefont{K.}~\bibnamefont{Orginos}},
  \bibinfo{author}{\bibfnamefont{V.}~\bibnamefont{Pascalutsa}},
  \bibnamefont{and}
  \bibinfo{author}{\bibfnamefont{M.}~\bibnamefont{Vanderhaeghen}},
  \bibinfo{journal}{PoS(LATTICE2008)}  (\bibinfo{year}{2008}),
  \eprint{arXiv.0809.1629}.

\bibitem[{\citenamefont{Lee et~al.}(2005)\citenamefont{Lee, Kelly, Zhou, and
  Wilcox}}]{Lee:2005ds}
\bibinfo{author}{\bibfnamefont{F.~X.} \bibnamefont{Lee}},
  \bibinfo{author}{\bibfnamefont{R.}~\bibnamefont{Kelly}},
  \bibinfo{author}{\bibfnamefont{L.}~\bibnamefont{Zhou}}, \bibnamefont{and}
  \bibinfo{author}{\bibfnamefont{W.}~\bibnamefont{Wilcox}},
  \bibinfo{journal}{Phys. Lett.} \textbf{\bibinfo{volume}{B627}},
  \bibinfo{pages}{71} (\bibinfo{year}{2005}), \eprint{hep-lat/0509067}.

\bibitem[{\citenamefont{Burkardt}(2000)}]{Burkardt:2000za}
\bibinfo{author}{\bibfnamefont{M.}~\bibnamefont{Burkardt}},
  \bibinfo{journal}{Phys. Rev.} \textbf{\bibinfo{volume}{D62}},
  \bibinfo{pages}{071503} (\bibinfo{year}{2000}).

\bibitem[{\citenamefont{Miller}(2007)}]{Miller:2007uy}
\bibinfo{author}{\bibfnamefont{G.~A.} \bibnamefont{Miller}},
  \bibinfo{journal}{Phys. Rev. Lett.} \textbf{\bibinfo{volume}{99}},
  \bibinfo{pages}{112001} (\bibinfo{year}{2007}).

\bibitem[{\citenamefont{Carlson and
  Vanderhaeghen}(2008{\natexlab{a}})}]{Carlson:2007xd}
\bibinfo{author}{\bibfnamefont{C.~E.} \bibnamefont{Carlson}} \bibnamefont{and}
  \bibinfo{author}{\bibfnamefont{M.}~\bibnamefont{Vanderhaeghen}},
  \bibinfo{journal}{Phys. Rev. Lett.} \textbf{\bibinfo{volume}{100}},
  \bibinfo{pages}{032004} (\bibinfo{year}{2008}{\natexlab{a}}).

\bibitem[{\citenamefont{Carlson and
  Vanderhaeghen}(2008{\natexlab{b}})}]{Carlson:2008zc}
\bibinfo{author}{\bibfnamefont{C.~E.} \bibnamefont{Carlson}} \bibnamefont{and}
  \bibinfo{author}{\bibfnamefont{M.}~\bibnamefont{Vanderhaeghen}}
  (\bibinfo{year}{2008}{\natexlab{b}}), \eprint{arXiv.0807.4537}.

\bibitem[{\citenamefont{Deser et~al.}(2000)\citenamefont{Deser, Pascalutsa, and
  Waldron}}]{Deser:2000dz}
\bibinfo{author}{\bibfnamefont{S.}~\bibnamefont{Deser}},
  \bibinfo{author}{\bibfnamefont{V.}~\bibnamefont{Pascalutsa}},
  \bibnamefont{and} \bibinfo{author}{\bibfnamefont{A.}~\bibnamefont{Waldron}},
  \bibinfo{journal}{Phys. Rev.} \textbf{\bibinfo{volume}{D62}},
  \bibinfo{pages}{105031} (\bibinfo{year}{2000}).

\end{thebibliography}

\end{document}